\documentclass[12pt]{article}

\advance\voffset by -2.0cm \advance\hoffset by -1.25cm
\usepackage{amsmath}
\usepackage{amssymb}
\usepackage{amsthm}
\usepackage{amsfonts}

\theoremstyle{definition}

\newcommand{\RR}{\mathbb{R}} 

\def\l{\label}
\def\S{{\cal S}}

\newcommand{\be}{\begin{equation}}
\newcommand{\ee}{\end{equation}}

\newlength{\oldcolsep}

\begin{document}

\title{Superluminality and a Curious Phenomenon in the Relativistic Quantum Hamilton-Jacobi Equation}
\author{Marco Matone}\date{}

\maketitle

\begin{center}Dipartimento di Fisica ``G. Galilei'' and Istituto
Nazionale di Fisica Nucleare \\
Universit\`a di Padova, Via Marzolo, 8 -- 35131 Padova, Italy
\\ e-mail: matone@pd.infn.it Phone: +39 049 827 7142, Fax: +39 049 827 7102\end{center}

\bigskip

\centerline{\it Dedicated to Shawn Lane}

\bigskip

\begin{abstract}
A basic problem in the relativistic quantum Hamilton-Jacobi theory
is to understand whether it may admit superluminal solutions. Here
we consider the averaging of the speed on a period of the oscillating
term which is similar to Dirac's
averaging of the oscillating part of the free electron's speed.
Such an averaging solves the problem of getting the $\hbar\to 0$
limit of the speed of the free particle, and leads to solutions
that, depending on the integration constants, may be superluminal.
\end{abstract}

\newpage

A basic issue related to the relativistic quantum Hamilton-Jacobi
equation is to investigate its superluminal solutions.
This is a subtle question involving the quantum potential and the
problem of defining the classical limit in the case of a free
particle. The main idea here is to average on the oscillating part,
a procedure which is reminiscent of Dirac's averaging of the
oscillating part of the free electron's speed ${i\over2}c\hbar
{\dot\alpha_1^0}e^{-2iHt/\hbar}H^{-1}$ \cite{Dirac}. This suggests a
new view on the quantum Hamilton-Jacobi equations.

The quantum version of the non-relativistic and relativistic
Hamilton-Jacobi equations \cite{Holland} has been derived by first
principles in \cite{1,BFM,2}, strictly related to classical-quantum and Legendre dualities
\cite{Faraggi:1996rn,Matone:1995rx,Matone:1995jr,Matone:2002wh}. In \cite{1} it has been shown that
energy quantization follows as a theorem without using any axiomatic
interpretation of the wave function.

A basic outcome of the formulation proposed in \cite{1,BFM,2}, is
that the quantum potential, that plays the role of intrinsic energy,
is never trivial. This is true even in the case of the free particle
classically at rest.

Consider the Klein-Gordon equation in the stationary case \be
(-\hbar^2c^2\Delta+m^2c^4-E^2)\psi=0 \ . \l{zero}\ee The
relativistic stationary quantum Hamilton-Jacobi equation follows by
setting $\psi=Re^{{i\over \hbar}\S_0}$ \be (\nabla
\S_0)^2+m^2c^2-{E^2\over c^2}-\hbar^2{\Delta R\over R}= 0 \ ,
\label{uno}\ee where $\S_0$ and $R$ satisfy the continuity equation
\be \nabla\cdot(R^2\nabla \S_0)=0 \ . \label{continuity}\ee In terms
of the quantum potential
$$
Q=-{\hbar^2\over 2m}{\Delta R\over R} \ ,
$$
and of the conjugate momentum
$$p=\nabla \S_0 \ ,
$$ Eq.(\ref{uno}) reads
\be
E^2=p^2c^2+m^2c^4+2m Qc^2 \ .
\label{due}\ee
For our purpose it is sufficient to consider the $1+1$ dimensional case. The continuity equation gives $R=1/\sqrt{S_0'}$, so that
$$
Q={\hbar^2\over4m}\{S_0,q\} \ ,
$$
where $\{f,q\}={f'''\over f'}-{3\over2}\big({f''\over f'}\big)^2$ is the Schwarzian derivative of $f$.
Therefore
(\ref{uno}) and (\ref{continuity}) reduce to the single equation
\be
(\partial_q S_0)^2+m^2c^2-{E^2\over c^2}+{\hbar^2\over2}\{S_0,q\}=0 \ .
\label{unodim}\ee

\noindent Eq.(\ref{unodim}) is formally equivalent to the quantum Hamilton-Jacobi equation which follows from the
stationary Schr\"odinger equation, namely
\be
{1\over2m}(\partial_q S_0)^2+V-E+{\hbar^2\over4m}\{S_0,q\}=0 \ ,
\label{unodimschro}\ee
where $V(q)$ is the potential.
It follows that the general solution of (\ref{unodim}) has the same form of the one of (\ref{unodimschro}), that is \cite{1,BFM,2}
\be
e^{{2i\over\hbar}\S_0\{\delta\}}=e^{i\alpha}{w+i\bar\ell\over
w-i\ell} \ , \l{s}\ee where $w=\psi^D/\psi\in\RR$ with $\psi$ and
$\psi^D$ two real linearly independent solutions of the
Klein-Gordon equation (\ref{zero}) in the 1+1 dimensional case.
Furthermore, we have $\delta=\{\alpha,\ell\}$, with $\alpha\in\RR$
and $\ell=\ell_1+i\ell_2$ integration constants. The crucial point
here is to note that $\ell_1\ne 0$ even when $E=mc^2$, equivalent to
having $\S_0\ne cnst$, which is a necessary condition to define
$\{\S_0,q\}$. This implies a non-trivial $\S_0$, even for a particle
classically at rest.

Following Floyd \cite{Floyd}, time parametrization is defined by Jacobi's theorem
\be
t={\partial {\S_0}\over\partial E} \ ,
\l{tempo}\ee
that, since $p=\partial_q \S_0$, implies $v=\partial E/\partial p$, which is the group velocity.
It should be stressed that (\ref{tempo}) is different from the time parametrization which follows by the conventional Bohmian mechanics, where
$$
v={p\over m}\neq \Big({\partial^2 {\S_0}\over\partial q\partial E}\Big)^{-1} \ .
$$
Furthermore, the formulation of the quantum Hamilton-Jacobi equation derived in \cite{1,BFM} solves the problem, first observed by Einstein, of considering
the classical limit in the case of bound states, such as in the case of the harmonic oscillator.

Set $L=k\ell$, $L_1=\Re L$, $L_2=\Im L$ where
$$k={1\over \hbar c}\sqrt{E^2-m^2c^4} \ . $$
Note that $L$ may depend on the particle quantum numbers. Even if,
as shown in \cite{1,BFM,2}, $L$ may depend on the energy and
fundamental constants as well, here we consider the case of $L$
independent of $E$. Two linearly independent solutions of
(\ref{zero}) are $\sin (kq)$ and $\cos (kq)$, so that by (\ref{s})
we have that in the case of $\S_0$, solution of (\ref{unodim}), the
mean speed is \be v={q\over t}=c{\sqrt{E^2-m^2c^4}\over
E}H_E(L_1,L_2;q)  \ , \l{velocity}\ee with \be H_E(L_1,L_2;q)={
|\sin(kq)-i L \cos(kq)|^2\over L_1} \ . \l{H}\ee
We see that for $L=1$
$$
H_E(1,0;q)=1 \ ,
$$
so that one recovers the classical relativistic relation
$$
v=c{\sqrt{E^2-m^2c^4}\over E} \ .
$$
The situation changes considerably if $L_1\neq 1$ and/or $L$ has a
non-trivial imaginary part, $L_2\neq0$, even small. In this case,
due to the oscillating terms, the limit $\hbar\to0$ is not
well-defined. Such a problem also arises in the non-relativistic case. Actually, the above expressions can be obtained directly from the ones of free non-relativistic particle
replacing $E$ by ${E^2\over2mc^2}-{mc^2\over2}$. In \cite{1,BFM,2} it has been shown that taking the $\hbar\to0$ limit requires introducing fundamental constants, in particular the
Planck length. This is of particular interest since the analysis concerned the non-relativistic formulation of the quantum Hamilton-Jacobi equation and may be seen as a signal
that at the level of foundations of quantum mechanics, special relativity should be taken into account. A related issue is that in spite of the fact that the Heisenberg uncertainty principle is fundamental,
there is no a rigorous formulation of an analog relativistic principle leading to the uncertainty relations in the $c\to \infty$ limit. Its heuristic formulation follows by the observation that whether
$c$ is the maximal speed, then the relativistic version of the Heisenberg uncertainty relations should have an upper bound on the uncertainty of the momentum, namely $\Delta p\sim mc$. In particular  \cite{Landau}
$$
\Delta q\sim {\hbar\over mc} \ .
$$

We now show that there is a natural way of getting the $\hbar\to 0$ limit which interestingly leads to solutions which also admit superluminality.
First, one may easily check that
there are values of $q$, $L_1$ and $L_2$ for which \be
H_E(L_1,L_2;q)>1 \ , \l{nice}\ee which is a basic hint that
superluminal solutions exist. Furthermore, we note that there is a natural way to get rid of the
oscillatory terms which opens a new view on the quantum
Hamilton-Jacobi theory. In particular, averaging on the short period
leads to an expression which is independent of $\hbar$, so providing an alternative to the
undefined $\hbar\to0$ limit. Let us restrict to the case $|L|=1$, that is
set \be L_1=\cos\theta \ , \qquad L_2=\sin\theta \ , \ee so that
Eq.(\ref{velocity}) reduces to \be v={q\over
t}=c{\sqrt{E^2-m^2c^4}\over E}{1+\sin\theta \, \sin(2kq)\over
\cos\theta} \ . \l{velocityddd}\ee Note that the term $\sin(2kq)$ is
very rapidly oscillating (e.g., in the case of \cite{OPERA}, the
neutrino Compton wavelength $\lambda_\nu$ is approximately $10^{11}$
times $k^{-1}$). This justifies averaging $v$ on the interval $[q,
q+{\pi\over k}]$, so that Eq.(\ref{velocityddd}) becomes \be \langle
v\rangle ={k\over \pi}\int^{q+{\pi\over k}}_qv(q')dq'=
c{\sqrt{E^2-m^2c^4}\over E}{1\over \cos\theta} \ .
\l{velocityatjpi}\ee This is the analog of Dirac's averaging
of the oscillating part of the free electron's speed \cite{Dirac}.
By (\ref{velocityatjpi}) we see that, depending on the values of
$\theta$, $E$ and $m$, one may have $\langle v\rangle>c$.

Another interesting consequence of (\ref{velocityatjpi}) is that,
since it corresponds to (\ref{velocityddd}) at $q=j\pi/2k$, it may
indicates a discrete space-time. In this respect, it would be very
interesting to formulate the relativistic quantum Hamilton-Jacobi
equation on a space-time lattice and check if it may lead to an
analogue of Eq.(\ref{velocityatjpi}).

\vspace{.333cm}

\noindent {\bf Acknowledgements}. It is a pleasure to heartily thank A.E. Faraggi for the collaboration over the years.

\newpage

\bibliography{apssamp}

\end{document}